\documentclass{article}

\usepackage{arxiv}

\usepackage{amsmath, amssymb, amsthm}
\usepackage{algorithm} 
\usepackage{algpseudocode}
\usepackage{tikz}
\usetikzlibrary{arrows,shapes.geometric,positioning,calc}
\usepackage[colorinlistoftodos]{todonotes}
\usepackage{changepage}
\usepackage{subfig}
\usepackage{graphicx}
\usepackage{soul}
\usepackage{todonotes} 
\usepackage{enumerate}
\usepackage{etextools}
\usepackage{mathtools}
\usepackage{listings}
\usepackage{xspace}
\usepackage{xcolor}

\newtheorem{theorem}{Theorem}[section]
 
\definecolor{codegreen}{rgb}{0,0.6,0}
\definecolor{codegray}{rgb}{0.5,0.5,0.5}
\definecolor{codepurple}{rgb}{0.58,0,0.82}
\definecolor{backcolour}{rgb}{0.95,0.95,0.92}
 
\lstdefinestyle{mystyle}{
    backgroundcolor=\color{backcolour},   
    commentstyle=\color{codegreen},
    keywordstyle=\color{magenta},
    numberstyle=\tiny\color{codegray},
    stringstyle=\color{codepurple},
    basicstyle=\ttfamily\footnotesize,
    breakatwhitespace=false,         
    breaklines=true,                 
    captionpos=b,                    
    keepspaces=true,                 
    numbers=left,                    
    numbersep=5pt,                  
    showspaces=false,                
    showstringspaces=false,
    showtabs=false,                  
    tabsize=2
}
 
\newcommand{\al}{\textit{et al.}\xspace}

\newcommand{\R}[0]{\mathbb{R}}

\newcommand{\Z}[0]{\mathbb{Z}}

\usepackage{todonotes}

\AtBeginDocument{%
  \providecommand\BibTeX{{%
    \normalfont B\kern-0.5em{\scshape i\kern-0.25em b}\kern-0.8em\TeX}}}

\title{The cross cyclomatic complexity:\\ a bi-dimensional measure for program complexity on graphs}

\author{Hugo Tremblay\\
Université du Québec à Chicoutimi\\
Chicoutimi, Québec\\
\texttt{Hugo\_Tremblay2@uqac.ca}\\
\And
Fabio Petrillo\\
Université du Québec à Chicoutimi\\
Chicoutimi, Québec\\
\texttt{fabio@petrillo.com}\\
}

\begin{document}
\maketitle

\begin{abstract}
Reduce and control complexity is an essential practice in software design. Cyclomatic complexity (CC) is one of the most popular software metrics, applied for more than 40 years. Despite CC is an interesting metric to highlight the number of branches in a program, it clearly not sufficient to represent the complexity in a piece of software. 
In this paper, we introduce the cross cyclomatic complexity (CCC), a new bi-dimensional complexity measure on graphs that combines the cyclomatic complexity and the weight of a minimum-weight cycle basis in as pair on the Cartesian plan to characterize program complexity using control flow graphs.
Our postulates open a new venue to represent program complexity, and we discuss its implications and opportunities.
\end{abstract}


\section{Introduction}

Maintainability is one of an essential quality in a software system; maintain a program drives evolution and the longevity of a system. Thus, to maintain a piece of software, firstly, we need to understand it. Understandability is the quality aspect that determines the capacity to understand code easily. However, how to measure if our source code is easy to understand? Despite more than 40 years of research for measuring understandability, there is not a consensus on what metrics we should use to evaluate the understandability \cite{Vinju2012}.

As understandability is hard to measure, the software engineering community uses the McCabe’s Cyclomatic Complexity \cite{McCabe1976} as a proxy of understandability. McCabe’s Cyclomatic Complexity (MCC) measures the number of linearly independent control flow paths within the method, claiming explicitly that \textit{"... it  can be used to manage and control program complexity."} \cite{McCabe1976} There is a commonsense and intuition if a program has more branches,  effort to understand it is more significant than a program with fewer branches. Then, if a method has a high CC, intuitively, it should be hard to understand.  

Despite its value, MCC is not enough to describe the complexity of a problem. For example, idiomatic Java structures as blocks \textit{try/catch} do not contribute to increase a program MCC. Moreover, a well-design program - using structure style - can have the same McCabe’s Cyclomatic Complexity than a equivalent program written using GOTOs or "spaghetti" code \cite{Vinju2012}. Further, Jbara \al found hundreds of methods with MCC higher than 100
\footnote{Visser \al propose good software system should have the majority of its methods with a MCC lower or equal 10. Then, a method with MCC of 100 should be considered as very complex.}, but they appreared to be well structured \cite{Jbara2014}. In fact, MCC does not reflect the complexity in all diversity of situation in a code, espcially in high-MCC programs \cite{Jbara2014}.

In this position paper, we introduce the \textbf{cross cyclomatic complexity (CCC)}, a new \textbf{bi-dimensional} complexity measure on graphs that combines the cyclomatic complexity and the weight of a minimum-weight cycle basis as a pair on the Cartesian plan to characterize program complexity using control flow graphs.

The remainder of this paper is organized as follows.
In the Section \ref{sec:mcc}, we discuss the characteristics and the main issues of MCC. In Section \ref{sec:math}, the mathematical and graph background to define our metric. In Section \ref{sec:ccc}, we define CCC metric, and Section \ref{sec:computing}, we present how to calculate CCC from 6 different cases. 
In the Section \ref{sec:relatedwork} we present related work to this study. Section \ref{sec:discussion} discusses the implications of CCC and opportunities of future work.

\section{Is McCabe's Cyclomatic Complexity a bad measure?}
\label{sec:mcc}

Cyclomatic Complexity (MCC) is a measure of program complexity that was proposed by Thomas McCabe in 1976. MCC measures the complexity of a program counting the number of linearly independent paths in a control flow graph \cite{McCabe1976}. Explicitly, McCabe's measure goal is \textit{"... provide a quantitative basis for modularization and allow us to identify software modules that will be difficult to test or maintain."} Indeed, MCC is a straightforward metric in terms of computation on which developers have an upper bound for the number of test cases required to obtain branch coverage of the code, and cyclomatic complexity could be applied to estimate the required effort for writing tests \cite{Hummel2014}. Also, MCC has strong mathematical support on graph theory, differently of arbitrary measures as Cognitive Complexity \cite{Campbell2016}, for example. Moever, a recent study found that there is no strong linear correlation between MCC and SLOC of Java methods, so developers should not use only SLOC as a proxy of complexity measure. Thus, it is not a surprise that MCC is one of the most adopted program metrics partly because it is simple to calculate, aligned with developers intuition and there are no widely agreed alternatives \cite{Ajami2019}. 

Despite some advantages, MCC has very well-know limitations \cite{Ajami2019, Jbara2014, Hummel2014, Campbell2016, Vinju2012}. First, intuitively \textit{while} loops are feels more complicate than \textit{if}s, as nested sequences of \textit{for}s and \textit{if} are more complicated than a \textit{switch/case} structure. However, MCC can give the same value because it gives the same weight to all structures. We can observe this issue in the Listing \ref{lst:sameccc}. The method \textit{sumOfPrimes} is intuitively more complex (lines 1 to 12) than the method getWords (lines 14 to 23). The first method has 2 \textit{for} loops and a nested \textit{if} inside of the second for, and a \textit{continue} to jump outside of the loops (as GOTO). The second method has just a \textit{switch/case} with 3 alternatives. Despite the intuitive differences in terms of complexity, both methods have the same MCC (MCC = 4). 

Moreover, a method with high MCC yet would easy to understand, as a large state machine implemented using a \textit{switch/case} structure. 
Thus, MCC may produce a false positive conclusion about the complexity of methods \cite{Vinju2012, Landman2014}. 
Finally, Ajami\al \cite{Ajami2019} found that different code structures take different times to interpret, suggesting that the approach taken by MCC where all branching constructs are given the same weight, is overly simplistic.

\begin{lstlisting}[language=Java, caption=Two Java methods with the same MCC \cite{Campbell2016}, label=lst:sameccc]
int sumOfPrimes(int max) {              
  int total = 0;                        
  OUT: for (int i = 1; i <= max; ++i) { 
    for (int j = 2; j < i; ++j) {       
      if (i % j == 0) {                 
        continue OUT;
      }
    }
    total += i;
  }
  return total;
}                  // Cyclomatic Complexity 4

String getWords(int number) {   
    switch (number) {           
      case 1:                   
        return "one";
      case 2:                   
        return "a couple";
      default:                  
        return "lots";
    }
  }        // Cyclomatic Complexity 4
\end{lstlisting}

Alternatives to MCC that use different weights to different control structures ware proposed (as Halstead metrics \cite{Halstead1977} or Cognitive Complexity \cite{Campbell2016}, but they did not report empirical evidence supporting proposed weights \cite{Ajami2019}. Thus, Hummel \cite{Hummel2014} suggest that 

\begin{quote}
\textit{The sad truth is that there probably is \textbf{no single simple measurement} that can express an abstract concept such as complexity in a single number. But this does not imply that we cannot measure and control complexity. It just has to be done with \textbf{multiple metrics and checks that cover the various aspects of complexity}.}
\end{quote}

\begin{figure}[h]
\centering
\includegraphics[width=1\linewidth]{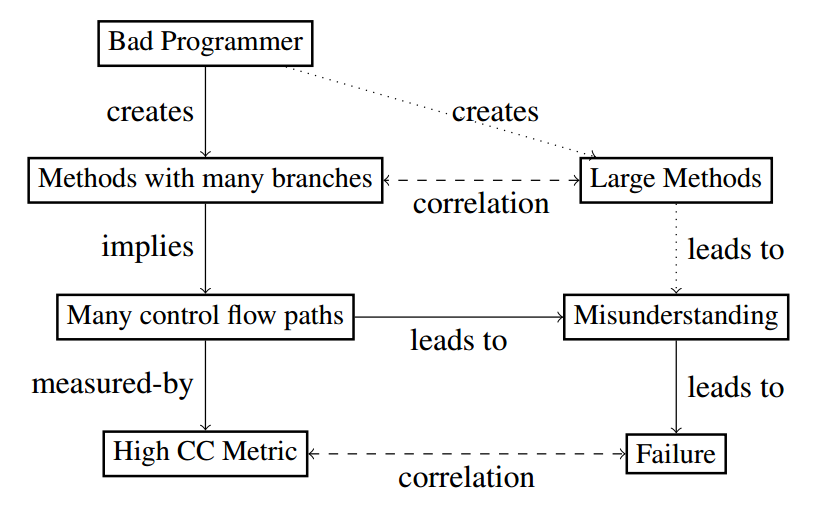}
\caption{Two comparable explanations for
correlation of high CC with failure \cite{Vinju2012}}
\label{fig:cc_with_failure}
\end{figure}

Vinju and Godfrey \cite{Vinju2012} share a similar perspective. They propose a schematic diagram (Figure \ref{fig:cc_with_failure}) to highlight and explain the factors for the correlation of high CC with failure. The diagram suggests the number of failures \textbf{should not be determinate by a single aspect} (number of branches). Thus, they highlighted the fact that there are several aspects to explain the failures. 

All in all, MCC is a useful measure to help developers to find methods that are hard to test, but MCC is not enough. Thus,  in Section \ref{sec:math}, we present essential concepts before introducing our metric. After, we introduce our metric in Section \ref{sec:ccc}.

\section{Mathematical prerequisites}
\label{sec:math}

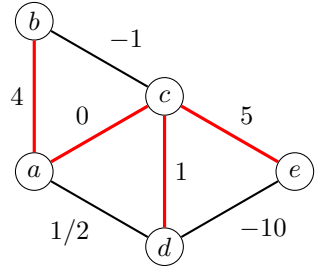
\begin{figure}
	\centering

    \begin{tikzpicture}[>=latex, scale=1, auto,
	sommet/.style={draw, circle, minimum size=0.5 cm, inner sep=0 cm}]

	\node[sommet] (a) at (0,0) {$a$};
	\node[sommet] (b) at ($(a)+(90:2)$) {$b$};
	\node[sommet] (c) at ($(a)+(30:2)$) {$c$};
	\node[sommet] (d) at ($(c)+(-90:2)$) {$d$};
	\node[sommet] (e) at ($(d)+(30:2)$) {$e$};
	
	\draw[red, very thick] (a) -- node[black] {$4$} (b);
	\draw[thick] (b) -- node[black] {$-1$} (c);
	\draw[red, very thick] (a) -- node[black] {$0$} (c);
	\draw[thick] (a) -- node[black, swap] {$1/2$} (d);
	\draw[red, very thick] (c) -- node[black] {$1$} (d);
	\draw[red, very thick] (c) -- node[black] {$5$} (e);
	\draw[thick] (d) -- node[black, swap] {$-10$} (e);
	
\end{tikzpicture}

	\caption{A weighted graph $G=(V,E_G)$ and one of its spanning tree $T=(V,E_T)$. Here, $\omega(G)=-1/2$ and $\omega(T)=10$.}
	\label{fig:graph}
\end{figure}

We now recall some important concepts from graph theory and linear algebra necessary in order to define our metric. Additional examples and details of proofs can be found in \cite{GrossYellen2005}. Let $G=(V,E)$ be a connected graph without loops or multiple edges. If the edges of $G$ are non-ordered pairs (resp. ordered pairs), we say $G$ is a simple graph (resp. a simple digraph). A $uv$-path is a sequence of adjacent edges $P(u,v)=(e_1,e_2,\ldots,e_n)$ such that $u$ is incident to $e_1$, $v$ is adjacent to $e_n$. A cycle $C(u,v)$ is a path where $u=v$. Also, $P(u,v)$ is simple if it does not contain repeated vertices, except for possibly $u$ and $v$.  A spanning tree of $G$ is an acyclic subgraph $T$ such that $V_T=V_G$. Spanning trees are easily computed using Breadth-First Search or Depth-First Search algorithms. Note that in the case of digraphs, those algorithms return a spanning tree only if $G$ is strongly connected, that is if there exists a path between each pair of vertices. Otherwise, the resulting tree depends on the choice of a starting vertex. A weighted graph is a graph $G=(V,E)$ together with a function $\omega:E\longrightarrow \R$ where the image $\omega(e)$ is the edge-weight of e. By extension, the weight of a subset $E'\subseteq E$ is the sum $\omega(E')=\sum_{e\in E'}\omega(e)$. The weight of a graph is the weight of its edge set. For example, if $\omega(e)=1$ for all $e\in E$, then $\omega(P)$ corresponds to the length of the path $P$. Figure \ref{fig:graph} illustrates a weighted graph and one of its spanning tree.

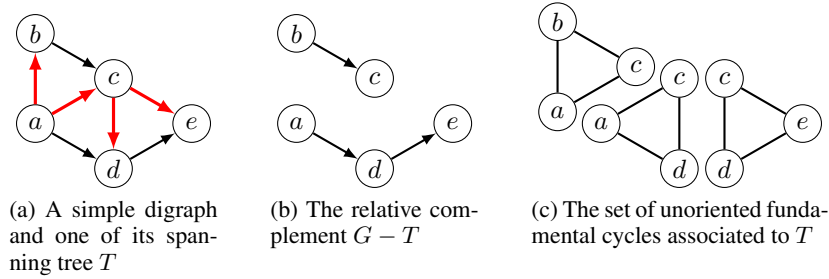
\begin{figure}
	\centering

	\subfloat[A simple digraph and one of its spanning tree $T$]{\label{fig:fundamental-cycles-a}
	\begin{tikzpicture}[>=latex, scale=0.6, auto,
	sommet/.style={draw, circle, minimum size=0.5 cm, inner sep=0 cm}]

	\node[sommet] (a) at (0,0) {$a$};
	\node[sommet] (b) at ($(a)+(90:2)$) {$b$};
	\node[sommet] (c) at ($(a)+(30:2)$) {$c$};
	\node[sommet] (d) at ($(c)+(-90:2)$) {$d$};
	\node[sommet] (e) at ($(d)+(30:2)$) {$e$};
	
	\draw[red, very thick, ->] (a) -- (b);
	\draw[thick, ->] (b) -- (c);
	\draw[red, very thick, ->] (a) -- (c);
	\draw[thick, ->] (a) -- (d);
	\draw[red, very thick, ->] (c) -- (d);
	\draw[red, very thick, ->] (c) -- (e);
	\draw[thick, ->] (d) -- (e);
	
\end{tikzpicture}
}\qquad
	\subfloat[The relative complement $G-T$]{\label{fig:fundamental-cycles-b}
	\begin{tikzpicture}[>=latex, scale=0.6, auto,
	sommet/.style={draw, circle, minimum size=0.5 cm, inner sep=0 cm}]

	\node[sommet] (a) at (0,0) {$a$};
	\node[sommet] (b) at ($(a)+(90:2)$) {$b$};
	\node[sommet] (c) at ($(a)+(30:2)$) {$c$};
	\node[sommet] (d) at ($(c)+(-90:2)$) {$d$};
	\node[sommet] (e) at ($(d)+(30:2)$) {$e$};
	
	\draw[thick, ->] (b) -- (c);
	\draw[thick, ->] (a) -- (d);
	\draw[thick, ->] (d) -- (e);
	
\end{tikzpicture}
}
\qquad
	\subfloat[The set of unoriented fundamental cycles associated to $T$]{\label{fig:fundamental-cycles-b}
	\begin{tikzpicture}[>=latex, scale=0.6, auto,
	sommet/.style={draw, circle, minimum size=0.5 cm, inner sep=0 cm}]

	\node[sommet] (a) at (0,0) {$a$};
	\node[sommet] (b) at ($(a)+(90:2)$) {$b$};
	\node[sommet] (c) at ($(a)+(30:2)$) {$c$};
	
	\draw[thick] (a) -- (b);
	\draw[thick] (b) -- (c);
	\draw[thick] (c) -- (a);
	
	\node[sommet] (a2) at ($(a)+(-15:1)$) {$a$};
	\node[sommet] (c2) at ($(a2)+(30:2)$) {$c$};
	\node[sommet] (d2) at ($(c2)+(-90:2)$) {$d$};
	
	\draw[thick] (a2) -- (c2);
	\draw[thick] (c2) -- (d2);
	\draw[thick] (d2) -- (a2);
	
	\node[sommet] (c3) at ($(c2)+(0:1)$) {$c$};
	\node[sommet] (d) at ($(c3)+(-90:2)$) {$d$};
	\node[sommet] (e) at ($(d)+(30:2)$) {$e$};
	
	\draw[thick] (c3) -- (d);
	\draw[thick] (d) -- (e);
	\draw[thick] (e) -- (c3);
	
\end{tikzpicture}
}
	\caption{Constructing a set of fundamental cycles}
	\label{fig:fundamental-cycles}
\end{figure}

Now, let $T$ be a spanning tree of $G$. The relative complement of $T$ in $G$ is the graph $G-T$ obtained from $G$ by deleting the edges of $T$. A fundamental cycle associated to $T$ is the cycle obtained by adding any edge of $G-T$ to $T$. Existence and unicity are guaranteed since $T$ is acyclic and spans $G$. The fundamental system of cycles of $T$ is the set of fundamental cycles associated to $T$. It is worth noting that in the case of simple digraphs, the cycles obtained are not oriented (see Figure \ref{fig:fundamental-cycles} for an example). This will be further discussed in the next section.

As its name suggest, a fundamental system of cycles depends on the choice of $T$. The following discussion on the links between linear algebra and graph theory expands on this idea. Recall from linear algebra that $GF(2)$ (often noted $\Z/2\Z$) is the Galois Field on $2$ elements. That is, the set $\{0,1\}$ with addition and multiplication modulo $2$. Now, let $E_1$ and $E_2$ be subsets of $E$ for a given graph $G=(V,E)$. Define the ring sum $E_1\oplus E_2$ as the symmetric difference between $E_1$ and $E_2$, that is $$E_1\oplus E_2 = (E_1-E_2)\cup (E_2-E_1).$$ If we denote by $W_E(G)$ the set of all subsets of $E_G$, then $W_E(G)$ together with $\oplus$ is a vector space over $GF(2)$. Finally, the cycle space of $G$ is the subspace of $W_E(G)$ consisting of $\emptyset$, all cycles of $G$ and all union of disjoint cycles. The next theorem establishes important results concerning basis of the cycle space.

\begin{theorem}\label{thm:basis}
    Let $G$ be a simple connected graph. Then, any fundamental system of cycles is a basis of the cycle space of $G$. Also, the dimension of the cycle space is equal to $\#E-\#V+1$. 
\end{theorem}

The dimension of the cycle space is called the cycle rank of $G$ (or Betti number or cyclomatic number) and is denoted by $\nu(G)$. For example, the graph of Figure \ref{fig:graph} has cycle rank $3$. The cycles $c_1=\langle a,e,b\rangle$, $c_2=\langle b,e,d\rangle$ and $c_3=\langle b,d,c\rangle$ form a basis and we have $c_1\oplus c_2=\langle a,e,d,b\rangle$.

By considering the control flow graph $G$ of a computer program and adding an extra virtual arc between the end and start node, one obtains the well-known cyclomatic complexity of the program, as first introduced by McCabe in \cite{McCabe1976}.

\begin{figure}
	\centering

	\subfloat[Basis associated to $T_1$: $\{\langle a,e,d\rangle, \langle a,d,c,b\rangle, \langle a,c,b\rangle\}$ of weight $36$]{\label{fig:weigth-cycle-basis-a}
\begin{tikzpicture}[>=latex, scale=1, auto,
	sommet/.style={draw, circle, minimum size=0.5 cm, inner sep=0 cm}]

	\node[sommet] (a) at (0,0) {$a$};
	\node[sommet] (b) at ($(a)+(-90:1.5)$) {$b$};
	\node[sommet] (c) at ($(b)+(180:2)$) {$c$};
	\node[sommet] (d) at ($(c)+(90:1.5)$) {$d$};
	\node[sommet] (e) at ($(d)+(0:1)+(90:1)$) {$e$};
	
	\draw[red, very thick] (a) -- node[black] {$1$} (b);
	\draw[thick] (a) -- node[black, swap] {$3$} (c);
	\draw[red, very thick] (a) -- node[black, swap] {$5$} (d);
	\draw[red, very thick] (a) -- node[black, swap] {$4$} (e);
	\draw[red, very thick] (b) -- node[black] {$2$} (c);
	\draw[thick] (c) -- node[black] {$7$} (d);
	\draw[thick] (d) -- node[black] {$6$} (e);
	
\end{tikzpicture}
}\qquad
	\subfloat[Basis associated to $T_2$: $\{\langle a,e,d\rangle, \langle a,d,c,b\rangle, \langle a,d,c\rangle\}$ of weight $45$]{\label{fig:weigth-cycle-basis-b}
\begin{tikzpicture}[>=latex, scale=1, auto,
	sommet/.style={draw, circle, minimum size=0.5 cm, inner sep=0 cm}]

	\node[sommet] (a) at (0,0) {$a$};
	\node[sommet] (b) at ($(a)+(-90:1.5)$) {$b$};
	\node[sommet] (c) at ($(b)+(180:2)$) {$c$};
	\node[sommet] (d) at ($(c)+(90:1.5)$) {$d$};
	\node[sommet] (e) at ($(d)+(0:1)+(90:1)$) {$e$};
	
	\draw[thick] (a) -- node[black] {$1$} (b);
	\draw[thick] (a) -- node[black, swap] {$3$} (c);
	\draw[red, very thick] (a) -- node[black, swap] {$5$} (d);
	\draw[thick] (a) -- node[black, swap] {$4$} (e);
	\draw[red, very thick] (b) -- node[black] {$2$} (c);
	\draw[red, very thick] (c) -- node[black] {$7$} (d);
	\draw[red, very thick] (d) -- node[black] {$6$} (e);
	
\end{tikzpicture}
}
\qquad
	\subfloat[Basis associated to $T_3$: $\{\langle a,e,d\rangle, \langle a,d,c\rangle, \langle a,c,b\rangle\}$ of weight $36$]{\label{fig:weigth-cycle-basis-c}
\begin{tikzpicture}[>=latex, scale=1, auto,
	sommet/.style={draw, circle, minimum size=0.5 cm, inner sep=0 cm}]

	\node[sommet] (a) at (0,0) {$a$};
	\node[sommet] (b) at ($(a)+(-90:1.5)$) {$b$};
	\node[sommet] (c) at ($(b)+(180:2)$) {$c$};
	\node[sommet] (d) at ($(c)+(90:1.5)$) {$d$};
	\node[sommet] (e) at ($(d)+(0:1)+(90:1)$) {$e$};
	
	\draw[red, very thick] (a) -- node[black] {$1$} (b);
	\draw[red, very thick] (a) -- node[black, swap] {$3$} (c);
	\draw[red, very thick] (a) -- node[black, swap] {$5$} (d);
	\draw[red, very thick] (a) -- node[black, swap] {$4$} (e);
	\draw[thick] (b) -- node[black] {$2$} (c);
	\draw[thick] (c) -- node[black] {$7$} (d);
	\draw[thick] (d) -- node[black] {$6$} (e);
	
\end{tikzpicture}
}
	\caption{Different cycle basis and their weigth}
	\label{fig:weigth-cycle-basis}
\end{figure}
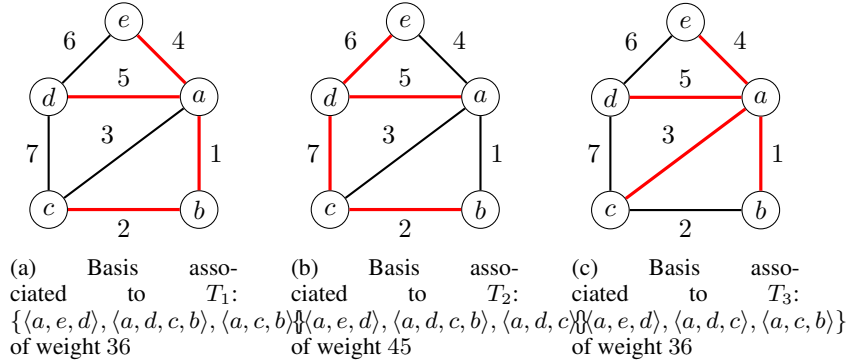

We close this section by studying cycle basis for weighted graphs. Given a weighted graph $G=(V,E)$, the weight of a basis is the sum of the weight of all its cycles. A minimum-weight basis is a cycle basis such that its weight is smaller than any other basis. Figure \ref{fig:weigth-cycle-basis} shows three possible cycle basis for a weigthed graph.

\section{The cross cyclomatic complexity}
\label{sec:ccc}

As stated in Section \ref{sec:mcc}, MCC is an interesting metric to measure complexity of graphs but is clearly not sufficient. Indeed, many structural properties of $G$ are osculated by $\nu(G)$. Thus, we could augment cyclomatic complexity by considering, on one axis, $\nu(G)$, and on another, the weight of a minimum-weight cycle basis for $G$. Intuitively, it is an extension of the traditional MCC: {\bf Whereas $\nu(G)$ merely gives the number of independent cycles in $G$, $\omega_{min}(G)$ describes the internal structure of those cycles}. Also, as is the case for the $\nu(G)$, the weight of a minimum-weight cycle basis is well-defined for any given graph. So, we define a new bi-dimensional complexity measure on graphs and study its applications to control flow graphs. Let $G=(V,E)$ be a weighted simple graph or digraph. Then, the \textbf{cross cyclomatic complexity}\footnote{Technically, since no cross product is involved, it should be called something along the line of "cartesian cyclomatic complexity". We chose the term "cross" simply because we found it slightly less boring.} of G is the tuple

\begin{equation}
c_\omega(G)=\left(\nu(G), \omega_{min}(G)\right).
\end{equation}

where $\omega_{min}(G)$ is the weight of a minimum-weight cycle basis for $G$.

We now describe how $c_\omega$ can be applied to measure complexity of computer programs. A control flow graph is a simple digraph $G=(V,E)$ where all paths are oriented from a start vertex $s$ to an exit vertex $r$ together with a virtual arc $(r,s)$. Such graphs are always strongly connected, otherwise some part of the program would be inaccessible. It is worth noting that cycles in $G$ correspond to independent paths in $G-\{(r,s)\}$. We define the cross cyclomatic complexity of a computer program as $c_\omega$ with $\omega(e)=1$ for all $e\in G$. It is obvious from the definition that $\nu(G)=\#E-\#V+2$. The computation of $c\omega(G)$ is a little more involved and thus is studied in the next section. Finally, our metric offers some flexibility on how to treat independent paths in control flow graphs. Indeed, \textbf{by assigning different weight to arcs, one could, for instance, penalize certain paths in the control flow graph}. Remark that we insist on using both $\nu(G)$ and $\omega_{min}(G)$ in the expression for $c_\omega$ because, although $\omega_{min}$ comes from $\nu$, one cannot explicitly compute $\nu$ from the value of $\omega_{min}$.

\begin{figure}
	\centering

\begin{tikzpicture}[polyomino/.style={black, very thick}, >=latex, scale=1.2,
	cell/.style={draw, thick, rectangle, minimum width=1cm, minimum height=1cm, fill=black, fill opacity=0.4, shift={(0.5,0.5)}}]
	
	\draw[help lines, opacity=0.7] (-0.1,-0.1) grid (5.1, 5.1);
	
	\draw[thick,->] (0,0) -- (0, 5.3);
	\draw[thick,->] (0,0) -- (5.8, 0);
	
	\node at (5.5,-0.2) {$\nu$};
	\node at (-0.5,5.3) {$\omega_{min}$};
	
	\draw[very thick, opacity = 0.4] (0,0) -- (5.5,5.5);
	
	\draw[fill, fill opacity=0.2, opacity= 0.2] (0,0) -- (5.5,0) -- (5.5,5.5) -- (0,0);
	
	\fill (0,0) circle (2pt);
	\fill[blue] (1,1) circle (3pt);
	\node at (1,1.3) {(a)};
	\fill (1,2) circle (2pt);
	\fill (1,3) circle (2pt);
	\fill (1,4) circle (2pt);
	\fill (1,5) circle (2pt);
	
	\fill[blue] (2,3) circle (3pt);
	\node at (2,3.3) {(b)};
	\fill[blue] (2,4) circle (3pt);
	\node at (2,4.3) {(c) and (d)};
	\fill (2,5) circle (2pt);
	
	\fill (3,4) circle (2pt);
	\fill (3,5) circle (2pt);
	
	\fill (4,5) circle (2pt);
	
	\node[below] at (1,0) {$1$};
	\node[left] at (0,1) {$1$};

\end{tikzpicture}

	\caption{The cross complexity halfplane. Dots correspond to possible values for the metric. Blue dots correspond to the complexity of the control structures in Figure \ref{fig:atomic}}
	\label{fig:plane}
\end{figure}
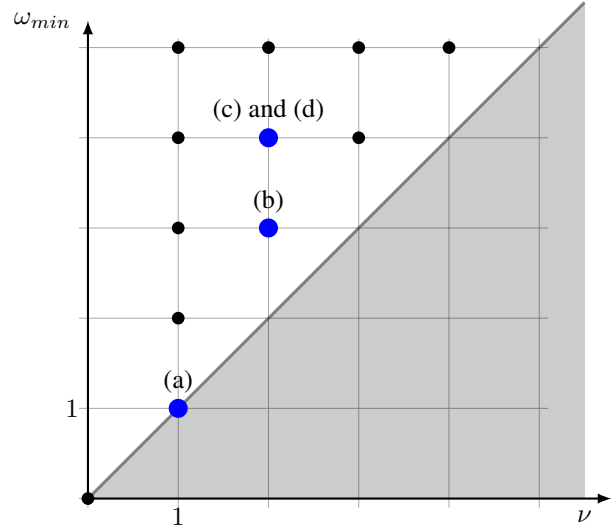

Before closing this section, we derive some useful properties of $c\omega$. First, since $\omega(e)=1$ for all $e\in G$, $\omega_{min}(G)$ corresponds to the length of a minimal-length cycle basis. Also, $\omega_{min}(G)\geq \nu(G)$ since any cycle contains at least one edge. Further, the equality holds if and only if $G$ is an isolated vertex or a sequence of two instructions. Finally, we can plot the complexity in a halfplane as in Figure \ref{fig:plane}. Section \ref{sec:computing} presents examples supporting this claim.

\section{Computing the CCC for control flow graphs}
\label{sec:computing}

As mentioned in Section \ref{sec:math}, computing a minimum-weight cycle basis is not trivial. Several algorithms have been proposed \cite{BergerGritzmannDeVries2009,Horton1987}.

We present here a $\mathcal{O}(\#E^3\#V)$ algorithm due to Horton \cite{Horton1987} (Algorithm \ref{algo:Horton}). The idea is to construct a $\mathcal{O}(\#E\#V)$-sized set of cycles guaranteed to contain a minimum-weight cycle basis. Then, elements of this set are greedily extracted and linear independence in checked using Gaussian elimination.

\begin{algorithm}
\caption{Horton} \label{algo:Horton}
  \begin{algorithmic}[1] 
  \Require{A simple weighted digraph $G=(V,E)$}
  \Ensure{A minimum-weight cycle basis $\mathcal{B}$ for $G$}
    \State{$\mathcal{B}\gets \emptyset$, $L\gets \emptyset$}
	\For{$x,y\in V$}
	    \State{Find the shortest $xy$-path $P(x,y)$} \label{line:Dijkstra}
	\EndFor
	 \For{$e=(x,y)\in E$ and $z\in V$}
	    \State{Add $C(z,e)=P(z,x)+e+P(y,z)$ to $L$ if it is simple}
	\EndFor
	\State{Order $L$ by cycle weight}
	\While{$\#\mathcal{B}<\nu(G)$}
	\State $c\gets$ the first element of $L$
	\If{the elements of $\mathcal{B}$ and $c$ are linearly independent} \label{line:Gauss}
	    \State Add $c$ to $\mathcal{B}$
	\EndIf
	\State Remove $c$ from $L$
	\EndWhile
	\State \Return $\mathcal{B}$
  \end{algorithmic}
\end{algorithm}

Line \ref{line:Dijkstra} can be computed using, for example, Dijkstra's algorithm \cite{Dijkstra1959}. For Line \ref{line:Gauss}, one can check check linear independence as follows: Given a cycle $c=\langle e_1,e_2,\ldots, e_n\rangle$, construct the $\#E$-dimensional incidence vector $[\delta_{e_i}]_{e_i\in E}$ where $\delta_{e_i}=1$ if and only if $e_i$ belongs to $c$ and $0$ otherwise. Then, linear independence of a set of cycle $\{c_1,c_2,\ldots, c_k\}$ is easily checked by constructing the matrix $[c_1\; c_2\; \cdots c_k]$ and using Gaussian elimination over $GF(2)$.

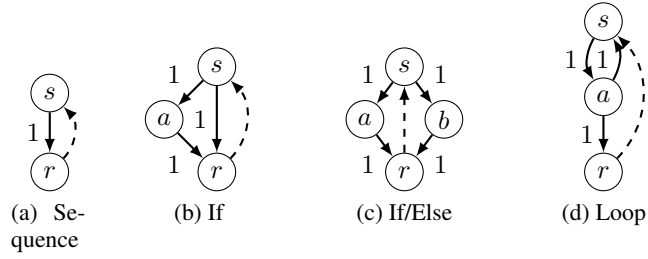
\begin{figure}[h]
	\centering

	\subfloat[Sequence]{\label{fig:atomic-a}
\begin{tikzpicture}[>=latex, scale=0.7, auto,
	sommet/.style={draw, circle, minimum size=0.5 cm, inner sep=0 cm}]

	\node[sommet] (a) at (0,0) {$s$};
	\node[sommet] (b) at ($(a)+(-90:1.5)$) {$r$};
	
	\draw[thick, ->] (a) -- node[black, swap] {$1$} (b);
	\draw[thick, dashed, ->] (b) edge[bend right=45] (a);
	
\end{tikzpicture}
}\qquad
	\subfloat[If]{\label{fig:atomic-b}
\begin{tikzpicture}[>=latex, scale=0.7, auto,
	sommet/.style={draw, circle, minimum size=0.5 cm, inner sep=0 pt}]

	\node[sommet] (a) at (0,0) {$s$};
	\node[sommet] (b) at ($(a)+(-90:1)+(180:1)$) {$a$};
	\node[sommet] (c) at ($(a)+(-90:2)$) {$r$};
	
	\draw[thick, ->] (a) -- node[black,swap] {$1$} (b);
	\draw[thick, ->] (a) -- node[black, swap] {$1$} (c);
	\draw[thick, ->] (b) -- node[black,swap] {$1$} (c);
	\draw[thick, dashed, ->] (c) edge[bend right=45] (a);
	
\end{tikzpicture}
}
\hskip 30pt
	\subfloat[If/Else]{\label{fig:atomic-c}
\begin{tikzpicture}[>=latex, scale=0.7, auto,
	sommet/.style={draw, circle, minimum size=0.5 cm, inner sep=0 pt}]

	\node[sommet] (a) at (0,0) {$s$};
	\node[sommet] (b) at ($(a)+(-90:1)+(180:0.75)$) {$a$};
	\node[sommet] (c) at ($(a)+(-90:1)+(0:0.75)$) {$b$};
	\node[sommet] (d) at ($(a)+(-90:2)$) {$r$};
	
	\draw[thick, ->] (a) -- node[black,swap] {$1$} (b);
	\draw[thick, ->] (a) -- node[black] {$1$} (c);
	\draw[thick, ->] (b) -- node[black,swap] {$1$} (d);
	\draw[thick, ->] (c) -- node[black] {$1$} (d);
	\draw[thick, dashed, ->] (d) -- (a);
	
\end{tikzpicture}
}
\hskip 30pt
	\subfloat[Loop]{\label{fig:atomic-d}
\begin{tikzpicture}[>=latex, scale=1, auto,
	sommet/.style={draw, circle, minimum size=0.5 cm, inner sep=0 pt}]

	\node[sommet] (a) at (0,0) {$s$};
	\node[sommet] (b) at ($(a)+(-90:1)$) {$a$};
	\node[sommet] (c) at ($(b)+(-90:1)$) {$r$};
	
	\draw[thick, ->] (a) edge[bend right] node[black,swap] {$1$} (b);
	\draw[thick, ->] (b) edge[bend right] node[black] {$1$} (a);
	\draw[thick, ->] (b) -- node[black,swap] {$1$} (c);
	\draw[thick, dashed, ->] (c) edge[bend right=45] (a);
	
\end{tikzpicture}
}
	\caption{The control structures of a program have cross cyclomatic complexity (a) $c_{min}=(1,1)$, (b) $c_{min}=(2,3)$, (c) $c_{min}=(2,4)$ and (d) $c_{min}=(2,4)$}
	\label{fig:atomic}
\end{figure}

For small graphs, the cross cyclomatic complexity of the control structures of a program are rather easy to compute since there exists very few cycle basis (see Figure \ref{fig:atomic}). For instance, the unique (and thus minimal-weight) cycle basis for the graph in Figure \ref{fig:atomic-c} is the set of two independent cycles $\langle s,a,r\rangle$ and $\langle s,b,r\rangle$ giving complexity $c_{min}(G)=(2,4)$

A more complex example is presented in Figure \ref{fig:weigth-cycle-basis}. Using Horton's algorithm, one determine that the three lowest weight simple cycles are $c_1=\langle a,b,c\rangle$, $c_2=\langle a,e,d\rangle$ and $c_3=\langle a,d,c\rangle$ of weight $6$, $15$ and $15$ respectively. Since they are linearly independent, we find $\omega_{min}(G)=36$.

The associated matrix, with edge ordering $e_1$ through $e_7$, is $$\begin{pmatrix}1 &0 &0 \\ 1 &0 &0 \\ 1 &0 &1 \\ 0 &1 &0 \\ 0 &1 &1 \\ 0 &1 &0 \\ 0 &0 &1\end{pmatrix}.$$
It is easily checked that this matrix reduces to $$\begin{pmatrix}1 &0 &0 \\ 0 &1 &0 \\ 0 &0 &1 \\ 0 &0 &0 \\ 0 &0 &0 \\ 0 &0 &0 \\ 0 &0 &0\end{pmatrix},$$ and thus $c_1$, $c_2$ and $c_3$ are linearly independent. 

Despite being mathematically more accurate, the computation of $\omega_{min}(G)$ demands a significant effort. To  simplify its computation, we now propose a simplified version of our complexity measure. The idea rises from the following simple observation: Given a graph $G$ and one of its cycle basis $\mathcal{B}$, then \begin{equation}\omega_{min}(G)\leq \omega(\mathcal{B}).\end{equation}

From Section \ref{sec:math}, one can deduced the following simple algorithm for computing an upper bound on $\omega_{min}$.

\begin{algorithm}
\caption{Upper bound on $\omega_{min}$} \label{algo:approximation}
  \begin{algorithmic}[1] 
  \Require{A simple weighted digraph $G=(V,E)$}
  \Ensure{A value $\omega$ such that $\omega_{min}(G)\leq \omega$}
    \State{$\omega\gets 0$}
    \State{Construct a spanning tree $T$ of $G$}
    \For{$e\in G-T$}
        \State{$C(T,e)\gets$ the unique non-oriented cycle of $(G-T)\cup{\{e\}}$}
        \State{$\omega\gets \omega+\omega(C(T,e))$}
    \EndFor
    \State \Return $\omega$
  \end{algorithmic}
\end{algorithm}

\begin{figure}[b]
	\centering

\begin{tikzpicture}[>=latex, scale=1.2, auto,
	sommet/.style={draw, circle, minimum size=0.5 cm, inner sep=2pt}]
	
	\def\length{1}

	\node[sommet] (s1) at (0,0) {};
	\node[sommet] (s2) at ($(s1)+(220:\length)$) {};
	\node[sommet] (s3) at ($(s1)+(-40:\length)$) {};
	\node[sommet] (s4) at ($(s3)+(240:\length)$) {};
	\node[sommet] (s5) at ($(s3)+(-60:\length)$) {};
	\node[sommet] (s6) at ($(s3)+(-90:2*\length)$) {};
	\node[sommet] (s7) at ($(s6)+(240:1*\length)$) {};
	
	\draw[thick, ->, red] (s1) -- (s2);
	\draw[thick, ->, red] (s1) -- (s3);
	\draw[thick, ->] (s3) -- (s2);
	\draw[thick, ->] (s2) -- (s4);
	\draw[thick, ->, red] (s3) -- (s5);
	\draw[thick, ->, red] (s3) -- (s4);
	\draw[thick, ->] (s4) -- (s5);
	\draw[thick, ->, red] (s5) -- (s6);
	\draw[thick, ->] (s4) -- (s6);
	\draw[thick, ->, red] (s6) -- (s7);
	\draw[thick, ->] (s2) -- (s7);

\end{tikzpicture}

	\caption{Control flow graph and spanning tree for graph $G1$ in \cite{McCabe1976} with $c_\omega(G1)=(6,24)$}.
	\label{fig:small-graph}
\end{figure}
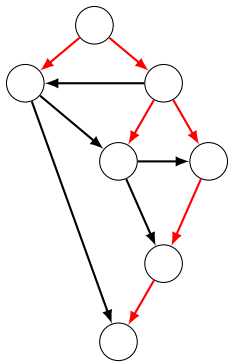

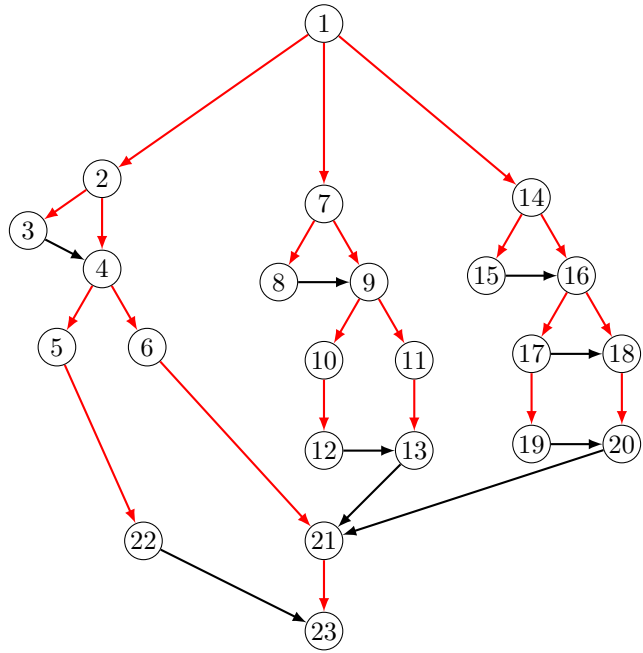
\begin{figure}
	\centering

\begin{tikzpicture}[>=latex, scale=1.2, auto,
	sommet/.style={draw, circle, minimum size=0.5 cm, inner sep=0 cm}]
	
	\def\length{1}

	\node[sommet] (s1) at (0,0) {$1$};
	\node[sommet] (s2) at ($(s)+(215:3*\length)$) {$2$};
	\node[sommet] (s3) at ($(s2)+(215:\length)$) {$3$};
	\node[sommet] (s4) at ($(s2)+(-90:\length)$) {$4$};
	\node[sommet] (s5) at ($(s4)+(240:\length)$) {$5$};
	\node[sommet] (s6) at ($(s4)+(-60:\length)$) {$6$};
	\node[sommet] (s7) at ($(s1)+(-90:2*\length)$) {$7$};
	\node[sommet] (s8) at ($(s7)+(240:\length)$) {$8$};
	\node[sommet] (s9) at ($(s7)+(-60:\length)$) {$9$};
	\node[sommet] (s10) at ($(s9)+(240:\length)$) {$10$};
	\node[sommet] (s11) at ($(s9)+(-60:\length)$) {$11$};
	\node[sommet] (s12) at ($(s10)+(-90:\length)$) {$12$};
	\node[sommet] (s13) at ($(s11)+(-90:\length)$) {$13$};
	\node[sommet] (s14) at ($(s1)+(-40:3*\length)$) {$14$};
	\node[sommet] (s15) at ($(s14)+(240:\length)$) {$15$};
	\node[sommet] (s16) at ($(s14)+(-60:\length)$) {$16$};
	\node[sommet] (s17) at ($(s16)+(240:\length)$) {$17$};
	\node[sommet] (s18) at ($(s16)+(-60:\length)$) {$18$};
	\node[sommet] (s19) at ($(s17)+(-90:\length)$) {$19$};
	\node[sommet] (s20) at ($(s18)+(-90:\length)$) {$20$};
	\node[sommet] (s21) at ($(s12)+(-90:\length)$) {$21$};
	\node[sommet] (s22) at ($(s21)+(180:2*\length)$) {$22$};
	\node[sommet] (s23) at ($(s21)+(-90:\length)$) {$23$};

	\draw[thick, ->, red] (s1) -- (s2);
	\draw[thick, ->, red] (s2) -- (s3);
	\draw[thick, ->, red] (s2) -- (s4);
	\draw[thick, ->] (s3) -- (s4);
	\draw[thick, ->, red] (s4) -- (s5);
	\draw[thick, ->, red] (s4) -- (s6);
	\draw[thick, ->, red] (s1) -- (s7);
	\draw[thick, ->, red] (s7) -- (s8);
	\draw[thick, ->, red] (s7) -- (s9);
	\draw[thick, ->] (s8) -- (s9);
	\draw[thick, ->, red] (s9) -- (s10);
	\draw[thick, ->, red] (s9) -- (s11);
	\draw[thick, ->, red] (s10) -- (s12);
	\draw[thick, ->, red] (s11) -- (s13);
	\draw[thick, ->] (s12) -- (s13);
	\draw[thick, ->, red] (s1) -- (s14);
	\draw[thick, ->, red] (s14) -- (s15);
	\draw[thick, ->, red] (s14) -- (s16);
	\draw[thick, ->] (s15) -- (s16);
	\draw[thick, ->, red] (s16) -- (s17);
	\draw[thick, ->, red] (s16) -- (s18);
	\draw[thick, ->] (s17) -- (s18);
	\draw[thick, ->, red] (s17) -- (s19);
	\draw[thick, ->, red] (s18) -- (s20);
	\draw[thick, ->] (s19) -- (s20);
	\draw[thick, ->, red] (s5) -- (s22);
	\draw[thick, ->, red] (s6) -- (s21);
	\draw[thick, ->] (s13) -- (s21);
	\draw[thick, ->] (s20) -- (s21);
	\draw[thick, ->, red] (s21) -- (s23);
	\draw[thick, ->] (s22) -- (s23);
	
\end{tikzpicture}

	\caption{Control flow graph and spanning tree for a graph in \cite{McCabe1976} with $c_\omega(G1)=(10,47)$.}
	\label{fig:big-graph}
\end{figure}

For example, the graph in Figure \ref{fig:weigth-cycle-basis-b} gives the approximation $(3,45)$ whereas $c_{\omega}=(3,36)$. In a more tangible example, consider the control flow graph of a Java implementation of Bubble sort from Rosetta Code \cite{rosettacode2020} (Figure \ref{fig:bubble sort}). Our approximation for the given spanning tree (red arcs) gives $c_\omega(G)=(4,12)$. Next, we extract from McCabe's original paper two more graphs for which he illustrated his measure (Figures \ref{fig:small-graph} and \ref{fig:big-graph}). The values $c_\omega(G)$ of each graph is shown in Figure \ref{fig:complexity}.

\begin{figure}
	\centering

\begin{tikzpicture}[>=latex, scale=1.2, auto,
	sommet/.style={draw, circle, minimum size=0.6 cm, inner sep=0 cm}]
	
	\def\length{1.2}

	\node[sommet] (s) at (0,0) {$s$};
	\node[sommet] (a) at ($(s)+(-90:\length)$) {$a$};
	\node[sommet] (b) at ($(s)+(-90:2*\length)$) {$b$};
	\node[sommet] (c) at ($(s)+(-90:3*\length)$) {$c$};
	\node[sommet] (d) at ($(s)+(-90:4*\length)$) {$d$};
	\node[sommet] (e) at ($(s)+(-90:5*\length)$) {$e$};
	\node[sommet] (r) at ($(s)+(-90:6*\length)$) {$r$};
	
	\draw[thick, ->, red] (s) -- (a);
	\draw[thick, ->, red] (a) -- (b);
	\draw[thick, ->, red] (b) -- (c);
	\draw[thick, ->, red] (c) -- (d);
	\draw[thick, ->, red] (e) -- (r);
	
	\draw[thick, dashed, ->] (r) edge[bend right=45] (s);
	\draw[thick, ->] (e) edge[bend right=45] (a);
	\draw[thick, ->] (d) edge[bend right=45] (b);
	\draw[thick, ->] (c) edge[bend right] (b);
	\draw[thick, ->, red] (b) edge[bend right] (e);
	
\end{tikzpicture}

	\caption{Control flow graph for Bubble Sort}
	\label{fig:bubble sort}
\end{figure}
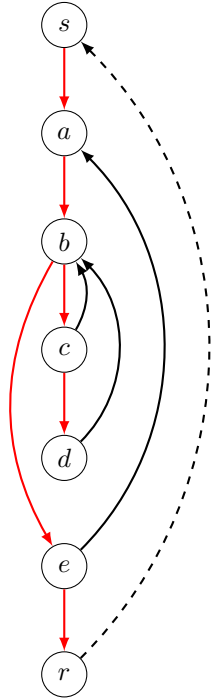

\begin{figure}
	\centering
\begin{tikzpicture}[polyomino/.style={black, very thick}, >=latex, scale=0.15,
	cell/.style={draw, thick, rectangle, minimum width=1cm, minimum height=1cm, fill=black, fill opacity=0.4, shift={(0.5,0.5)}}]
	
	\draw[help lines, opacity=0.7] (-0.1,-0.1) grid (50.1, 50.1);
	
	\draw[thick,->] (0,0) -- (0, 50.3);
	\draw[thick,->] (0,0) -- (50.8, 0);
	
	\node at (50.5,-2) {$\nu$};
	\node at (-5,50.3) {$\omega_{min}$};
	
	\draw[very thick, opacity = 0.4] (0,0) -- (50.5,50.5);
	\draw[very thick, opacity = 1] (0,0) -- (25.5,51);
	
	\draw[fill, fill opacity=0.2, opacity= 0.2] (0,0) -- (50.5,50.5) -- (25.5,50.5) -- (0,0);
	
	\draw[fill, fill opacity=0.4, opacity= 0.4] (0,0) -- (50.5,0) -- (50.5,50.5) -- (0,0);
	
	\fill[blue] (4,12) circle (15pt);
	\node at (4,15) {Fig \ref{fig:bubble sort}: $(4,12)$};
	
	\fill[blue] (10,47) circle (15pt);
	\node at (10,50) {Fig \ref{fig:big-graph}: $(10,47)$};
	
	\fill[blue] (6,24) circle (15pt);
	\node at (6,27) {Fig \ref{fig:small-graph}: $(6,24)$};
	
	\node[below] at (1,0) {$1$};
	\node[left] at (0,1) {$1$};

\end{tikzpicture}
	\caption{Cross cyclomatic complexity for a few computer programs}
	\label{fig:complexity}
\end{figure}
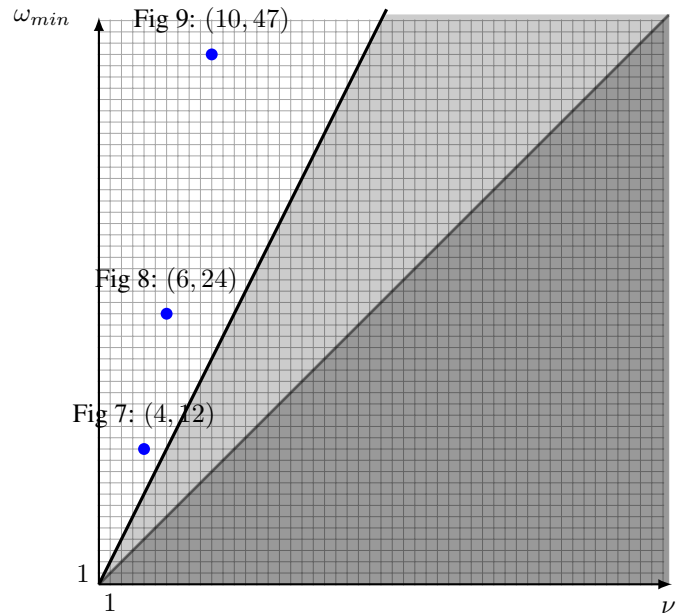

Analyzing Figure \ref{fig:complexity}, we can easily compare the three graphs' complexities. The three regions in the chart represent some properties of $c_\omega(G)$: No graph $G$ can have cross cyclomatic complexity located in the dark gray region of the plane since each cycle in a basis is of length at least $1$. The light gray region of the plane contains the complexity of valid graphs, but not non-trivial computer programs. Finally, the white region represents the $c_\omega(G)$ of valid non-trivial computer programs. In terms of design optimization, developers should aim to design programs with complexity near the line $\nu(G) = 2\omega_{min}$. Following this criterion, we observe that the Bubble sort implementation as modelled by the graph in Figure \ref{fig:bubble sort} is considered \textit{optimized} (near to the line), whereas the graph in Figure \ref{fig:big-graph} could be be refactored. 

\section{Related work}
\label{sec:relatedwork}

Vinju \al \cite{Vinju2012} investigated the relationship between the shape of control flow patterns observed in Java methods to their CC metric values, proposing the concept of abstract control flow patterns (CFPs) and compressed control flow patterns (CCFPs), which allow us to produce statistical evidence that the CC metric indeed does not adequately model the likely complexity of control flow in Java methods.

Jbara \al \cite{Jbara2014} discuss there are no agreement on how to measure code complexity, favouring simple general purpose metrics, such as lines of code or MCC. They claims such metrics just count syntactic features, and ignore details of the code’s global structure, which may also have an effect on understandability. To address this issue, they suggested that code regularity—where the same structures are repeated time after time—may significantly reduce complexity.

\section{Final remarks}
\label{sec:discussion}

Understanding a computer program is a time consuming and complex cognitive task. To control program complexity, researchers and practitioners have proposed single metrics to address this issue. However, those single metrics can not solve the problem yet. McCabe’s Cyclomatic Complexity is one of those metrics. The MCC abstracts the control flow graph as merely the analysis of branches and the fan-outs of its nodes. Doing so, it reduces the rich structure of the underlying graphs as a single number \cite{Vinju2012}.  

Thus, we claim for addressing the program complexity without throw out previous useful metrics (as MCC), we introduce the \textbf{cross cyclomatic complexity (CCC)}, a new \textbf{bi-dimensional} complexity measure on graphs that combines the cyclomatic complexity and the weight of a minimum-weight cycle basis as a pair on the Cartesian plane to characterize program complexity using control flow graphs. Using solid mathematical and graph backgrounds, CCC provides a combined approach to address some of the issues on MCC without losing its advantages.

Hence, we claim that Cross Cyclomatic Complexity \textbf{gives a criterion for highlight refactoring opportunities to control the complexity of computer programs. Namely, we suggest that one should aim to keep the complexity as close as possible to the line $\nu=\omega_{min}$}. Further, CCC is agnostic of computer languages, and robust graph theorems support it. Also, CCC addresses distortions of MCC, which ignores the complexity associated with the size of programs or structures as long \textit{switch/case} statements.

However, it is a positional work, lacking an in-depth empirical evaluation to observe how CCC can address complexity issues in program comprehension and support refactoring tasks. In the future, we will evaluate our metric to identify the diversity of scenarios to apply it and results on a large dataset of programs for different languages. Thus, this paper is the first step in order to provide a new way to see software metrics using graph properties.

\bibliographystyle{unsrt}
\bibliography{references}

\end{document}